\documentclass[prb,twocolumn,amsmath,amssymb,floatfix,showpacs,superscriptaddress]{revtex4}

\usepackage{bm}

\usepackage{amssymb,amsmath}
\usepackage{graphics}
\usepackage{subfigure}
\usepackage{epstopdf}
\usepackage{epsfig}
\usepackage{color}
\usepackage{amsfonts}
\usepackage{bm}
\usepackage{amscd}
\usepackage{epsfig}
\usepackage{subfigure}
\usepackage{color}
\usepackage{color}
\definecolor{darkgray}{rgb}{0.3,0.3,0.3}
\definecolor{gray}{rgb}{0.5,0.5,0.5}
\definecolor{yellow}{rgb}{.4,.4,0}
\definecolor{orange}{rgb}{1,0.5,0}
\definecolor{darkgreen}{rgb}{0,0.5,0}
\definecolor{darkblue}{rgb}{0,0,1}
\definecolor{darkred}{rgb}{0.5,0,0}
\definecolor{purple}{rgb}{0.35,0,0.35}

\begin{document}
\title{Enhanced performance  of joint cooling  and   energy production }

\author{ O. Entin-Wohlman}

\affiliation{Raymond and Beverly Sackler School of Physics and Astronomy, Tel Aviv University, Tel Aviv 69978, Israel}

\affiliation{Physics Department, Ben Gurion University of the Negev,  Beer Sheva
84105, Israel}
\email{oraentin@bgu.ac.il}

\author{Y. Imry}

\affiliation{Department of Condensed Matter Physics, Weizmann Institute of Science, Rehovot 76100 Israel}
\author{A. Aharony}

\affiliation{Raymond and Beverly Sackler School of Physics and Astronomy, Tel Aviv University, Tel Aviv 69978, Israel}
\affiliation{Physics Department, Ben Gurion University of the Negev,  Beer Sheva
84105, Israel}

\date{\today}

\begin{abstract}

The efficiencies/coefficients of performance of three-terminal devices, comprising two electronic terminals and a thermal one (e.g.,  a boson bath)
are discussed. In particular, two procedures are analyzed.  (a) One of the electronic terminals is cooled by investing thermal power (from the thermal bath) {\em } and  electric power (from  voltage applied across the electronic junction); (b) The invested thermal power from the boson bath is exploited to cool one electronic terminal {\em and }  to produce electric power.   Rather surprisingly,  the coefficient of performance of (b) can be enhanced as compared to that of (a).

\end{abstract}


\maketitle

\section{Introduction}
\label{intro}

Achieving high efficiencies and coefficients of performance in thermoelectric nanodevices is one of the main goals of contemporary research  on these effects. It is well-known that a strong energy dependence of   the electronic transport is a prerequisite for efficient thermoelectric phenomena involving charge carriers, e.g. the Seebeck effect.
In a seminal paper Mahan and Sofo \cite{MS}
proposed (Kedem and Caplan \cite{Kedem} advanced related ideas in 1965)
that high values of the thermoelectric efficiency of a two-terminal electronic device are
obtained when the energy-dependent conductance
has a sharp structure 
away from the nearly-common chemical potentials of the leads.
Following this proposal there were  quite a few attempts to achieve
effectively narrow electronic bands, especially in nanostructures with transmission resonances
and where the enhanced scattering of phonons at interfaces also reduces the phononic heat
conductance. \cite{Ven} Examples are quantum-well superlattices and quantum wires, \cite{Hicks} and crystalline    arrays  of quantum dots.  \cite {Cai} Likewise, the possibility    to    approach  in nanostructures  the  limit  of  reversible processes,   for which the efficiency is  the highest,   has  been  pursued, \cite{Humphrey} as well as the maximal efficiency at finite output power (see e.g. Ref.  \onlinecite{Whitney} and references therein).

A different way to achieve strongly energy-dependent transport is to consider inelastic processes, in which the electrons interchange energy with a boson bath, e.g., photons or electron-hole excitations (see Fig. \ref{fig1}). The boson bath coupled to the electronic system represents the third terminal, making  the setup a three-terminal one.
In such devices, in addition to the normal thermoelectric effects in the two electronic terminals, there can arise thermoelectric phenomena due to the energy transfer between the thermal terminal and the electronic ones. Physically, this is because the energy exchange between the electronic and bosonic systems induces  electronic (charge and heat) currents.

Various
three-terminal thermoelectric devices and processes have been proposed: setups designed for
cooling-by-heating  (CBH) processes, for which the bias voltage is kept zero and one of the electronic terminals is cooled by the thermal bath; \cite{Cleuren}  three-terminal junctions based on molecular bridges, where the charge carriers interchange energy with the vibrations of the molecule forming the bridge;  \cite{EIA} rectification of thermal fluctuations in chaotic cavities; \cite{Sanchez} two-sites nanostructures based on the inelastic phonon-assisted hopping; \cite{JHJ,JHJ1} cooling a two-dimensional electron gas
(which plays the role of the thermal bath) at low temperatures by elastic electron transitions to and from the leads; \cite{Edwards,Prance}
quantum ratchet converting the nonequilibrium noise of a nearby quantum point contact to dc current; \cite{Khrapai}
carbon nanotubes designed to extract energy from a discrete local oscillator at ultralow temperatures; \cite{Zippilli}
junctions connected to several electronic terminals; \cite{Sivan,Brandner,Mazza} and
cooling the vibrational motion by charge current. \cite{arrachea}

For a two-terminal system
converting  thermal energy into work
or conversely
cooling one of the terminals by investing  work, the maximal efficiency is achieved in the  reversible Carnot
thermodynamic cycle. 
When there are more than two terminals, one may define various efficiencies (or equivalently, coefficients of performance) and explore their limits in a reversible process. In Sec. \ref{YI} we examine, on general thermodynamic grounds,  two cooling scenarios
feasible in a  three-terminal setup and analyze their efficiencies    for zero entropy production.   Section \ref{general}  defines the currents and the thermodynamic forces driving them, and uses those to re-express the coefficients of performance introduced in Sec. \ref{YI}. In these two sections we  examine the coefficients of performance achieved in the corresponding reversible processes. 
Things become even more exciting once a specific model is introduced (in Sec. \ref{model}), and the efficiency is analyzed allowing for `parasitic' processes  in the junction, represented in our case by (unavoidable) phonon heat conductances (see Sec. \ref{IR}). We find that the three-terminal setup has interesting features. (a) When used to cool one of the electronic terminals by investing work (or alternatively an electric power) and  heat from the boson bath, the working range of the device increases as the latter increases;  (b) Depending on phonon conductances, the coefficient of performance for joint cooling and power production may be enhanced compared to the situation where work is invested. We summarize our results in Sec. \ref{SUM}.

\section{General thermodynamic considerations}

\label{YI}

\begin{figure}[ht]
\centering
\includegraphics[width=2.in]{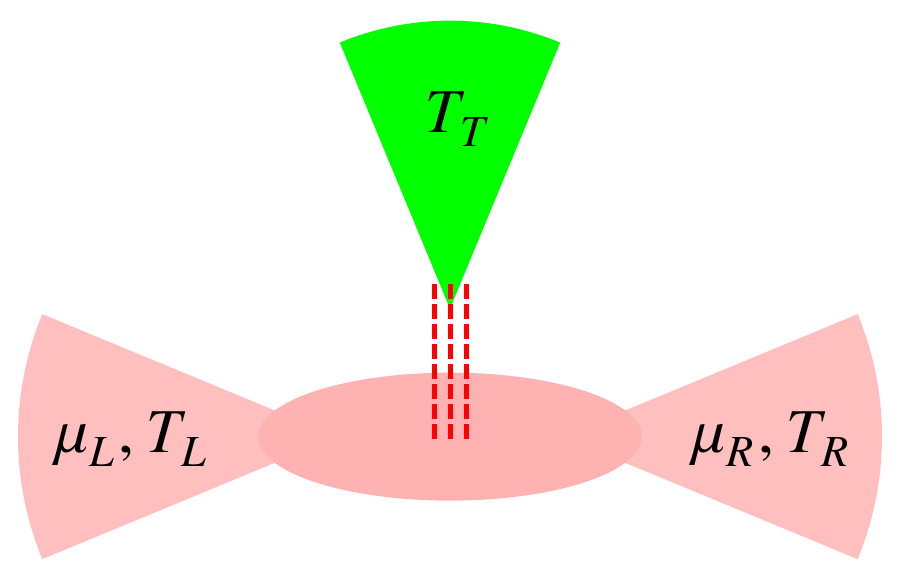}

\caption{(Color online)  The three-terminal setup. The electronic reservoirs on the left and on the right are characterized by their respective temperatures, $T_{L}$ and $T_{R}$, and chemical potentials, $\mu_{L}$ and $\mu_{R}$. The thermal (boson) terminal is held at a third temperature, $T_{T}$. The thin (red) lines indicate the bosons' flux exchanged between the electronic system and the thermal boson bath.}

 \label{fig1}
\end{figure}

The ubiquitous electronic thermoelectric nanodevice consists of a junction bridging  two  electronic terminals held at different temperatures and chemical potentials. We denote those by $ \mu_{L}$ and $T_{L}$ for the ``left" electronic terminal, and $ \mu_{R}$ and $T_{R}$ for the ``right" one.  Our three-terminal setup includes in addition  a thermal terminal
supplying bosons (e.g.,  phonons, photons, electron-hole excitations), thus allowing for inelastic transport processes of the charge carriers moving in-between the electronic terminals. 
The boson bath is kept at yet another temperature, denoted $T_{T}$ (see
Fig. \ref{fig1}).

We begin by examining  the three-terminal efficiency or  coefficient of performance (COP)   from a thermodynamic point of view, adopting a configuration similar to the cooling-by-heating (CBH) one. \cite{Cleuren}
W
e  choose to cool the left, $L$, terminal and dump the heat into the right, $R$, one, taking
\begin{align}
T^{}_L < T^{}_R\ .
\end{align}
Quite generally, a ``double-driving" can be exploited, comprising both invested work $W_{\rm in}$ (supplied for instance, by an electric current) 
and heat $Q_{T}$ (produced by the thermal reservoir at temperature $T_{T}$), to take heat $Q_{L}$ from the $L$ terminal and dump heat $-Q_{R}$ into the $R$ one. \cite{sign}
In the usual two-terminal thermoelectric cooling scenario $Q_{T}=0$ and the cooling is accomplished solely by the work $W_{\rm in}$.
The efficiency of this process when it is reversible, $\eta_{\rm cbe}$,  is
\begin{align}
\eta^{}_{\rm cbe}=T^{}_{L}/(T^{}_{R}-T^{}_{L})\ .
\label{etacbe}
\end{align}
The other particular scenario is when no work is invested, $W_{\rm in}=0$, and  cooling is achieved by $Q_{T}$, i.e., the CBH process, with the reversible efficiency $\eta_{\rm cbh}$
\begin{align}
\eta^{}_{\rm cbh}=\eta^{}_{\rm cbe}[1-(T_{R}/T_{T})]=\frac{1-(T_{R}^{}/T^{}_{T})}{(T^{}_{R}/T^{}_{L})-1}\ .
\label{etacbh}
\end{align}
To interpolate between these two limits we introduce the parameter
$\alpha = W_{\rm in}/Q_T$; in the first scenario $\alpha =\infty$ and in the second $\alpha =0$. In general,
$\alpha$ depends 
on the 
driving forces (see Secs. \ref{model} and \ref{IR}).

The cooling efficiency for the joint  process, i.e., the COP, is defined by
\begin{align}
\eta^{}_{a} = \frac{Q^{}_L}{W_{\rm in}^{} + Q^{}_T} = \frac{Q_L}{-Q_R - Q_L}\ ,
\label{eta}
\end{align}
where the second equality results from energy conservation
\begin{align}
-Q^{}_R = Q^{}_L + W^{}_{\rm in} +Q^{}_T\ .
\label{ec}
\end{align}
Since  the reservoirs are each in equilibrium, the corresponding entropies are
\begin{align}
S^{}_\ell = Q^{}_\ell/T^{}_\ell\ , ~~ (\ell= L,\ R,\ T)\ ,
\label{sd}
\end{align}
and the total entropy increment due to the process, 
$\Delta  S$, is
\begin{align}
\Delta S=S^{}_{L}+S^{}_{T}+S^{}_{R}\ .
\label{Ds}
\end{align}
In a reversible process $ \Delta S=0$; the corresponding COP, $\eta^{\rm rev}_{a}$, is  then [using Eqs. (\ref{ec}), (\ref{sd}),  and (\ref{Ds})]
\begin{align}
  \eta^{\rm rev}_{a} =\eta_{\rm cbe}  \Big [1 - \frac{T^{}_R}{T^{}_T (1+\alpha)}\Big ]\   .
 \label{3t eff}
\end{align}
As expected,
large values of  $\alpha$ give $\eta_{\rm cbe}$, Eq. (\ref{etacbe}),  while $\alpha = 0$ yields 
the maximal  efficiency of the cooling-by-heating process, Eq. (\ref{etacbh}).
Any finite positive $\alpha$ gives  $\eta^{\rm rev} _{a}< \eta_{\rm cbe}$, as follows trivially from Eq. (\ref{3t eff}). So, we lose efficiency by adding the CBH power. On the other hand, $\eta ^{\rm rev}_{a}> \eta_{\rm cbh}$, so we gain efficiency compared to  a pure CBH process.
The  advantages of the double driving are that we get more cooling power, that the working region of the device increases (see Secs. \ref{model} and \ref{IR})  and that for
realistic purposes, one could use the free sun as the thermal terminal,  \cite{Cleuren} which reduces the efficiency  by a few percent only.

Next we consider the case where  part of the invested heat $Q_{T}$ from the thermal terminal is exploited to cool the left electronic terminal and part of it is  converted to useful work, $W_{\rm out}$. Energy conservation in this case yields
 $Q_{T}+Q_{L}=W_{\rm out}-Q_{R}$, while the efficiency is
\begin{align}
\eta^{}_{b}=\frac{Q^{}_{L}+W^{}_{\rm out}}{Q^{}_{T}}\ .
\label{etab2}
\end{align}
For a reversible process,  the efficiency of the thermal terminal, which supplies both the useful work and the cooling of the left electronic terminal  is $1-T_{R}/T_{T}$, i.e.,
\begin{align}
Q^{}_{T}=\frac{W^{}_{\rm out}+W^{}_{\rm c}}{1-T^{}_{R}/T^{}_{T}}\ ,
\label{ref1}
\end{align}
where $W_{c}$ is the work used for cooling. 
As for a reversible process the efficiency of cooling is $\eta_{\rm cbe}$, Eq. (\ref{etacbe}),
it follows that
\begin{align}
Q^{}_{L}=\frac{W^{}_{\rm c}}{(T^{}_{R}/T^{}_{L})-1}\ .
\label{ref2}
\end{align}
Inserting \cite{referee}  the relations (\ref{ref1}) and (\ref{ref2}) into Eq. (\ref{etab2})
yields
\begin{align}
\eta^{\rm rev}_{b}=\Big (1-\frac{T^{}_{R}}{T^{}_{T}}\Big
) [w
+\eta^{}_{\rm cbe} (1-w)]\ ,
\end{align}
where $w$ is the fraction of the work produced in the process,
\begin{align}
w=\frac{W^{}_{\rm out}}{W^{}_{\rm out}+W^{}_{\rm c}}\ .\label{www}
\end{align}
Remarkably enough, the reversible efficiency increases with $w$ when $\eta_{\rm cbe}<1$ and decreases when $\eta_{\rm cbe}> 1$. We return to this point in Sec. \ref{general}. Like the ratio $\alpha=W_{\rm in}/Q_{T}$, $w$ also depends 
on the 
driving forces (see Secs. \ref{model} and \ref{IR}).


For completeness, we recast the above reasoning into the configuration in which the   thermal terminal is cooled by both work  done (e.g.,  by electrical current between the left and the right terminals, see Ref. \onlinecite{arrachea}) and a heat flow between  $L$ and  $R$, (here we take $T_T<T_R<T_L$).
The coefficient of performance for this process, [replacing Eq. (\ref{eta})] is
\begin{align}
\widetilde{\eta }= \frac{Q^{}_T}{W^{}_{\rm in} + Q^{}_L} = \frac{Q_T}{-Q^{}_R - Q^{}_T}\ ,
\label{eta'}
\end{align}
and  for a reversible process [i.e., $ \Delta S=0$, see Eq. (\ref{Ds})]
\begin{align}
\widetilde{ \eta}_{}^{\rm rev} = \eta' [1- \frac{T^{}_ R}{T^{}_L (1+\alpha ')}]\ , \label{3t eff'}
\end{align}
where $\alpha '= W_{\rm in}/Q_L$ and
$\eta'=T_{T}/(T_{R}-T_{T})$ is the usual Carnot efficiency  for cooling a terminal at temperature $T_{T}$  by moving heat  to a terminal having temperature $T_{R}$,
 using pure work.
When $\alpha ' \rightarrow \infty$ the reversible-process efficiency becomes simply $\eta'$, while for $\alpha ' \rightarrow 0$ the expression in the square brackets in Eq. (\ref{3t eff'}) becomes   $1-(T_{R}/T_{L})$. This is just the Carnot efficiency for converting the heat moved between the left and the right terminals to work. Equation (\ref{3t eff'}) is then the reversible efficiency for CBH, as found and explained by the first of Refs. \onlinecite{Cleuren} [see their Eq. (11)].

The above is very general, {\em beyond  linear-response transport}. It is straightforward to introduce the  (negative) correction to the efficiency
when the total entropy production (waste) is positive, as   
discussed in Sec. \ref{IR}.

\section{
currents and forces 
}

\label{general}

Having set the thermodynamic basis for defining  efficiencies (or coefficients of performance) in a three-terminal device,
we proceed to examine those from another point of view, by considering
the currents flowing  in the system and the thermodynamic forces driving them.  

The thermodynamic driving forces and the currents conjugate to them can be defined unambiguously by considering the entropy production of the device.
In terms of the heat/entropy production in the thermal terminal and in the electronic $L$ and $R$ terminals, $\dot{Q}_{T}$, $\dot{Q}_{L}$, and $\dot{Q}_{R}$ respectively, the entropy production is [{\it cf.}  Eqs. (\ref{sd}) and (\ref{Ds})]
\begin{align}
\Delta \dot{S}=\frac{\dot{Q}^{}_{T}}{T^{}_{T}}+\frac{\dot{Q}_{L}}{T^{}_{L}}+\frac{\dot{Q}^{}_{R}}{T^{}_{R}}\  .
\label{EP}
\end{align}
The rate of the  heat produced in each of the electronic reservoirs is
$\dot{Q}_{\ell}=\dot{E}_{\ell}-\mu_{\ell}\dot{N}_{\ell}$, $\ell=L $ or $R$ ($E_{\ell}$ is the  total  energy of the $\ell-$th electronic reservoir, and $N_{\ell}$ is the number of charge carriers there). Energy conservation implies that
\begin{align}
\dot{Q}_{T}+\dot{Q}^{}_{L}+\dot{Q}^{}_{R}=-\mu^{}_{L}\dot{N}^{}_{L}-\mu^{}_{R}\dot{N}^{}_{R}\ ,
\label{EC}
\end{align}
while charge conservation gives $\dot{N}_{L}+\dot{N}_{R}=0$.
Since the particle current emerging from the left electronic terminal is
\begin{align}
J^{}_{L}=-\dot{N}^{}_{L}\ ,
\label{JL}
\end{align}
it is seen that the right-hand side of Eq. (\ref{EC}) is just  the Joule heating in the system, $J_{L}(\mu_{L}-\mu_{R})$.

Exploiting the conservation laws,
the entropy production relation Eq. (\ref{EP})
becomes
\begin{align}
T^{}_{R}\Delta\dot{S}=J^{}_{L}(\mu^{}_{L}-\mu^{}_{R})+J^{Q}_{L}\Big (1-\frac{T^{}_{R}}{T^{}_{L}}\Big )+J^{Q}_{T}\Big (1-\frac{T^{}_{R}}{T^{}_{T}}\Big )\ .
\label{EP1}
\end{align}
Here we have chosen the temperature of the right electronic bath as the reference temperature, and introduced the  heat currents leaving the left terminal
\begin{align}
J^{Q}_{L}=-\dot{Q}^{}_{L}\ ,
\label{JLQ}
\end{align}
and the one leaving the thermal one
\begin{align}
J^{Q}_{T}=-\dot{Q}_{T}\ .
\label{JTQ}
\end{align}
As usual,  the entropy production Eq. (\ref{EP1})  appears as a (scalar) product of a vector consisting  of the three currents, $J^{}_{L}$, $J^{Q}_{L}$, and $J^{Q}_{T}$, with a vector comprising the driving forces,
the electric one,
\begin{align}
\mu^{}_{L}-\mu_{R}^{}=|e|V\ ,
\label{V}
\end{align}
and the two thermal  ones, given by the temperature difference across the electronic junction and  between the electronic system and the boson bath. We shall confine ourselves to the configuration where the left electronic terminal is to be cooled and will also use the thermal terminal as the chief power supplier. Then the lowest temperature of the three is $T_{L}$ and the highest one is $T_{T}$. Accordingly, the thermal driving forces are
$(T_{R}/T_{L})-1$ and $1-(T_{R}/T_{T})$.

One can imagine various  scenarios for cooling the $L-$electronic terminal in the three-terminal junction  depicted in Fig. \ref{fig1}. Here, as in Sec. \ref{YI}, we focus on two specific situations. We present for each of them the relevant coefficient of performance (COP) and then examine the best value it can have, which is obtained when the cooling and the accompanying processes are reversible, i.e., the entropy production vanishes.  The power(s) obtained  in such a process is (are) unfortunately vanishingly small. Nonetheless, it is instructive to investigate these upper limits that are the target of many studies in thermoelectricity (an example is the proposal of  Ref. \onlinecite{MS}). More realistic configurations and the corresponding coefficients of performance will be considered in Sec. \ref{IR}.

(a) As in Sec. \ref{YI},  the  $L-$electronic terminal can be cooled by investing  power extracted from the thermal terminal {\em and} an electric power. The 
COP  of this process, denoted $\eta_{a}$,  is the ratio of the thermal power gained and the two invested powers,  \cite{com1}
\begin{align}
\eta^{}_{a}=\frac{J^{Q}_{L}}{|e|J^{}_{L}V+J^{Q}_{T}}\ .
\label{etaa}
\end{align}
The  COP attains its  highest possible value in a reversible process, for which the entropy production vanishes. Indeed, upon using Eq. (\ref{EP1}) for $\Delta\dot{S}=0$
we find that in  a reversible process
\begin{align}
\eta^{\rm rev}_{a}=\frac{1}{(T^{}_{R}/T^{}_{L})-1}\times\frac{|e|J^{}_{L}V+[1-(T^{}_{R}/T^{}_{T})]J^{Q}_{T}}{|e|J^{}_{L}V+J^{Q}_{T}}\ .
\label{etaar}
\end{align}
When the electric power vanishes Eq. (\ref{etaar}) reproduces the COP of the `cooling-by-heating' process  in the reversible limit, \cite{Cleuren} given in Eq. (\ref{etacbh}), 
while in the more mundane scenario of cooling by investing only electric power,  the COP is $\eta_{\rm cbe}$  given by Eq. (\ref{etacbe}).

Obviously, when the entropy production  vanishes (or is rather small) the cooling-by-heating process is less effective than cooling by electric power; the  joint three-terminal COP lies in-between these two limits,
\begin{align}
\eta^{}_{\rm cbh}<\eta^{\rm rev}_{a}<\eta^{}_{\rm cbe}\ .
\label{ine}
\end{align}
The best performance
 in a reversible process is thus reached upon cooling by investing electric power. However, as compared to the reversible cooling-by-heating option proposed by Cleuren {\it et al.}  \cite{Cleuren} for which the electric voltage vanishes,  investing electric power in addition to the thermal one improves the COP. As mentioned, the three-terminal arrangement also extends the range of the `working condition' \cite{com1} of the device as compared to a two-terminal setup, see Sec. \ref{IR}.

(b) The left electronic terminal is cooled and at the same time electric power is being produced. Accordingly, the ratio between   the powers gained to that invested, i.e. the COP $\eta_{b}$, is
\begin{align}
\eta^{}_{ b}=\frac{J^{Q}_{L}+|eVJ^{}_{L}|}{J^{Q}_{T}}\ .
\label{etab}
\end{align}
When the process is reversible
the COP of this process is
\begin{align}
\eta^{\rm rev}_{b}=\Big (1-\frac{T^{}_{R}}{T^{}_{T}}\Big )\frac{J^{Q}_{L}+|eVJ^{}_{L}|}{J^{Q}_{L}[(T^{}_{R}/T^{}_{L})-1]
+|eVJ^{}_{L}|}\ .
\label{etabr}
\end{align}
We see that the reversible value of the COP when the voltage vanishes is given by  Eq. (\ref{etacbh}). On the other hand, if the electronic left bath is not cooled at all, then the device works as a `solar cell' (in the case where the thermal terminal is the sun), yielding   electric power in response to the temperature difference between the  electronic system and the thermal terminal. The best value of the COP for this configuration is the textbook Carnot efficiency $\eta_{\rm C}$
\begin{align}
\eta^{}_{\rm C}=1-T^{}_{R}/T^{}_{T}\ .
\label{Carnot}
\end{align}
As in case (a) above, $\eta^{\rm rev}_{b}$ is bounded in-between its  two limiting values. But in contrast to case (a),  which of the two is the bigger and which is the smaller depends on the temperature difference across the electronic reservoirs. When that temperature difference is significant,
$(T_{R}/T_{L})-1> 1$, i.e., $\eta_{\rm cbe}< 1$ [Eq. (\ref{etacbe})]
we find
\begin{align}
\eta^{}_{\rm C}>\eta^{\rm rev}_{b}> \eta^{}_{\rm cbh}\ ,\ \ \  \eta^{}_{\rm cbe}< 1\ .
\label{ine1}
\end{align}
When the temperature difference across the electrons is small then the inequality is reversed, i.e.,
\begin{align}
\eta^{}_{\rm cbh}>\eta^{\rm rev}_{b}> \eta^{}_{\rm C}\ ,\ \ \  \eta^{}_{\rm cbe}> 1\ .
\label{ine2}
\end{align}
At the crossing point, $\eta^{}_{\rm cbe}=1$,  $\eta_{\rm cbh}=\eta_{\rm C}$, and  the two bounds merge.
Examining Eqs. (\ref{ine1}) and (\ref{ine2}), we see that
when $\eta_{\rm cbe}<1$ it ``pays" to produce work (or power) in addition to cooling, while in the reverse case it does not.

 \section{ An example of a three-terminal thermoelectric device}

\label{model}

In order to consider the coefficients of performance away from reversibility one needs to invoke a specific system and find for it the currents flowing in the setup in response to the driving forces. In the linear-response regime the relation between  two vectors, the one of the currents and the one of the driving forces [{\it cf.} Eq. (\ref{EP1}) and the discussion around it] can be written in a matrix form
\begin{align}
\left [\begin{array}{c}|e|J^{}_{L} \\ \\ J^{Q}_{L}   \\ \\ J ^{Q}_{T} \end{array}\right ]=\frac{1}{T^{}_{R}}{\cal M}\left [\begin{array}{cc} V \\ \\
-[(T^{}_{R}/T^{}_{L})-1]\\ \\ 1-(T^{}_{R}/T^{}_{T})\end{array}\right ]\ ,
\label{ON}
\end{align}
where the (3$\times$3) matrix ${\cal M}$ comprises the transport coefficients. Within the linear-response regime ${\cal M}$
does not depend on the driving forces and is determined by the thermal-equilibrium
properties of the setup. It then obeys the Onsager relations; for instance, for a system invariant to time-reversal--the case studied here--${\cal M}$ is symmetric.
One notes that the matrix ${\cal M}$ is
nonnegative definite, to ensure the positiveness, or vanishing, of the entropy production $\Delta\dot{S}$,
Eq. (\ref{EP1}). The matrix ${\cal M}$ may  be singular, for example, when the  first     two rows are proportional to one another and then the two corresponding currents (for example, the electronic charge and thermal currents)  are  proportional to each other. This special situation, which was termed by
Kedem and Caplan\cite{Kedem} `strong coupling',  is also behind the mechanism of Mahan and Sofo, \cite{MS} involving transport in a very narrow  energy band.
This  yields high values of the COP. In the case of Ref. \onlinecite{MS}, each transferred electron carries the same energy and heat. We will come back to this point below. It can also happen that the third row is  also proportional   to the first one and then all the three currents, in the three-terminal case,  are proportional to each other, which may be termed `full strong  coupling'. The proportionality  coefficients are determined by the specific model at hand.

\begin{figure}[ht]
\centering
\includegraphics[width=2.2in]{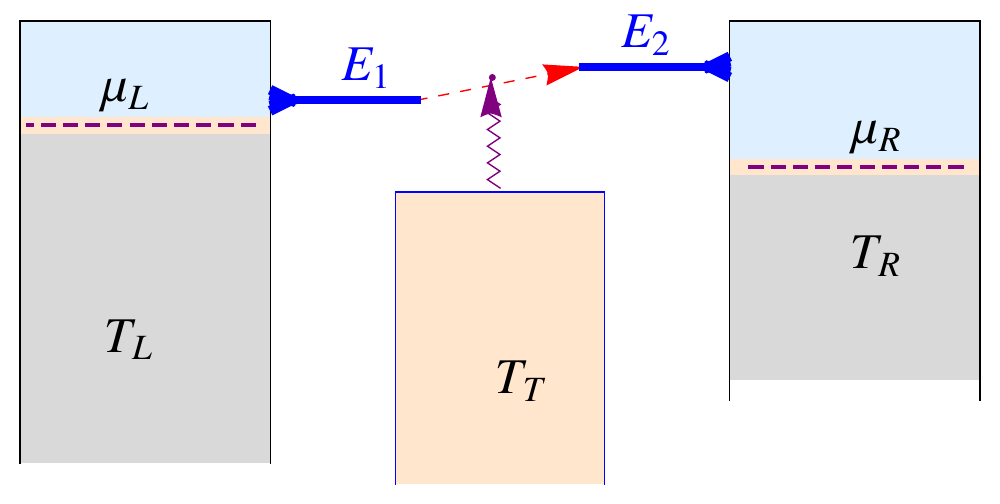}
\caption{ (Color online) Two electronic reservoirs are characterized by their respective electrochemical potentials, $\mu_{L}$ and $\mu_{R}$, and temperatures, $T_{L}$ and $T_{R}$.
The electronic transport between the two is accomplished  via
the two localized levels
of energies $E_{1}$ and $E_{2}$ that are well-coupled (elastically) each to its nearby  reservoir.
The left reservoir is cooled, i.e., $T_{L}<T_{R}$. The thermal reservoir (held at temperature $T_{T}$) supplies the required energy for the transport.
}
\label{fig2}
\end{figure}

We obtain the  transport coefficients, i.e.,   the matrix ${\cal M}$, for   the two-level model of Ref. \onlinecite{JHJ}, depicted in Fig. \ref{fig2}. The model exploited in Ref. \onlinecite{Cleuren} has two such two-level pairs, whose effects add in the cooling and subtract in the electrical current, causing the latter to vanish.
The setup displayed in Fig. \ref{fig2}
models a small one-dimensional nanosystem in which thermoelectric transport takes place mainly via inelastic phonon (or, more generally, boson)  assisted hopping, i.e., it is assumed that this inelastic hopping is the strongly-dominant electronic channel.
This will be the case for  temperatures above a certain threshold temperature, denoted in Ref. \onlinecite{JHJ} by $T_{x}$. Below that temperature
the transport is dominated by the elastic tunneling conductance. By equating the latter to the boson-assisted hopping conductance, one finds that $T_{x}$
is determined
by the energies $E_{1}$ and $E_{2}$
of the localized levels, times  the ratio of the localization length of the wave functions there to the junction linear dimension. \cite{JHJ}
 Any relevant inelastic transmission will be exponentially small when the temperature approaches zero.
As in previous publications on this issue \cite{Cleuren}
we discuss noninteracting quasiparticles. The main effect of the interactions is expected to be a renormalization of the model parameters, e.g., changing the energies $E_{1}$ and $E_{2}$ (see Fig. \ref{fig2})
as in the theory of Pollak \cite{Pollak} and Efros-Shklovskii \cite{ES} for the density of states.

In the two-level model of Ref. \onlinecite{JHJ} the transfer of the charge carriers in-between the left and right electronic terminals takes place via  two localized levels, of energies $E_{1}$ and $E_{2}$($>E_{1}$ for concreteness).
For instance, for an electron transferred from left to right, the boson bath gives an energy
$-E_{1}$ ($E_{2}$) to the left (right) lead
and thus the bosons transfer the energy
\begin{align}
\omega =E^{}_{2}-E^{}_{1}
\label{om}
\end{align}
to the electrons. A net energy of
\begin{align}
\overline{E}=\frac{1}{2}(E^{}_{1}+E^{}_{2})
\label{eb}
\end{align}
is transferred across the electronic system, from left to right.
It follows that (see Ref. \onlinecite{JHJ} for the details of the calculation), barring for the time-being any heat conduction by phonons present in the system, the electronic heat current is $J_{L}^{Q}=\overline{E}J^{}_{L}$ and the heat current flowing between the thermal bath and the electronic system is $J^{Q}_{T}=\omega J^{}_{L}$. The particle current $J_{L}$ is proportional  to the hopping conductance of the device.
Thus, when phonon conductances  are ignored, the setup is in the `full strong-coupling limit',  \cite{Kedem}  the 
matrix ${\cal M}$ is singular, and the entropy production vanishes.

In reality, the above picture has to be generalized by adding the elastic electronic transmission and, more importantly,
the phonon heat conductances: The heat current $J_{L}^{Q}$ should be augmented by the  phonons' heat flow between the two electronic terminals, and the current $J_{T}^{Q}$ by the phonons' flow between the boson bath and the electronic terminals.  \cite{com2}  We denote the phonon heat conductance  of the electronic system by $K_{\rm P}$, and between the electronic system and the thermal terminal by $K_{\rm PP}$. For simplicity, we confine ourselves to the regime where the elastic transport of the electrons may be ignored.  \cite{JHJ} Under these circumstances,
the matrix ${\cal M}$ giving the relations among the currents and the driving forces [see Eq. (\ref{ON})], 
is
\begin{align}
\frac{1}{T_{R}^{}}{\cal M}=G^{}_{\rm in}&\left[\begin{array}{ccc}1 \ \ &\overline{E}/|e|\ \ &\omega/|e| \\
\overline{E}/|e|\ \ &(\overline{E}/e)^{2}\ \ &\overline{E}\omega/e^{2} \\\omega/|e|\ \ &\overline{E}\omega/e^{2}\ \ & (\omega/e)^{2}\end{array}\right ]\nonumber\\
&+\left [\begin{array}{ccc}0\ \ &0\ \ &0 \\ 0\ \ & K^{}_{\rm P}\ \ &0 \\ 0\ \ &0\ \ &K^{}_{\rm PP}\end{array}\right ]\ .
\label{ONm}
\end{align}
Here $G_{\rm in} $ is the hopping conductance of the electronic junction.  The
  first term on the right-hand side of Eq. (\ref{ONm}) contains the strongly-coupled part of the transport matrix. The second term there, describing the parasitic phonon conductances $K^{}_{\rm P}$ and $K^{}_{\rm PP}$, 
  spoils this `strong coupling' property.

The working condition \cite{com1} for cooling the left electronic terminal requires that the heat current emerging from it be  positive. In our model, this amounts to
\begin{align}
&\frac{|e|V}{\overline{E}}-\Big (\frac{T^{}_{R}}{T^{}_{L}}-1\Big )+\frac{\omega }{\overline{E}}\Big (1-\frac{T^{}_{R}}{T^{}_{T}}\Big )
\geq \frac{e^{2}K^{}_{\rm P}}{\overline{E}^{2}G^{}_{\rm in}}\Big (\frac{T^{}_{R}}{T^{}_{L}}-1\Big )\ .
\label{wc}
\end{align}
Note that  the phonon thermal conductance on the right-hand side,  $K_{\rm P}$,  is scaled by the {\em    ``bare"} electronic thermal conductance of the junction, $G_{\rm in}\overline{E}^{2}/e^{2}$  [the transport coefficient contained in the   $22-$element of the matrix ${\cal M}$, see Eq. (\ref{ONm})].

It is illuminating to re-examine the entropy production Eq. (\ref{EP1})
in the framework of our model in conjunction with the working condition for cooling the left electronic terminal. Using Eq. (\ref{ONm}) in the relation (\ref{ON}),  and inserting the resulting currents into Eq.(\ref{EP1}) yields
\begin{align}
T^{}_{R}\Delta\dot{S}&=\frac{G^{}_{\rm in}}{e^{2}}\Big [|e|V-\overline{E}\Big (\frac{T^{}_{R}}{T^{}_{L}}-1\Big )+\omega \Big (1-\frac{T^{}_{R}}{T^{}_{T}}\Big )\Big ]^{2}\nonumber\\
&+K^{}_{\rm P}\Big (\frac{T^{}_{R}}{T^{}_{L}}-1\Big )^{2}+K^{}_{\rm PP}\Big (1-\frac{T^{}_{R}}{T^{}_{T}}\Big )^{2}\ .
\label{EPm}
\end{align}
As might have been expected, the ``parasitic" heat flows carried by the phonons [the last two terms in Eq. (\ref{EPm})] make the entropy production positive and the thermoelectric transport {\em irreversible.}
When these are ignored (or are very small) and our setup approaches the strong-coupling limit then the cooling process will be reversible provided that
$
|e|V/\overline{E}-[(T^{}_{R}/T^{}_{L})-1]+(\omega /\overline{E})[1-(T^{}_{R}/T^{}_{T})]=0$, i.e., when the heat flowing out of the left electronic terminal vanishes. In other words, when the cooling process is {\em reversible}  it yields, as is always the result,  {\em zero output power}.

\section{Irreversible coefficients of performance}

\label{IR}

Exploiting the transport coefficients of the specific device described in Sec. \ref{model},
the electric current leaving the left terminal is
\begin{align}
|e|J^{}_{L}=\frac{G^{}_{\rm in}\overline{E}}{|e|}\Big [\frac{|e|V}{\overline{E}}+1-
\frac{T^{}_{R}}{T^{}_{L}}+\frac{\omega}{\overline{E}}\Big (1-\frac{T^{}_{R}}{T^{}_{T}}\Big )\Big ]\ .
\end{align}
Likewise,
the electronic heat current from that terminal is
\begin{align}
J^{Q}_{L}=\frac{\overline{E}}{|e|}(|e|J^{}_{L})-
K^{}_{\rm P}\Big (\frac{T^{}_{R}}{T^{}_{L}}-1\Big )\ .
\end{align}
The working condition for the left terminal to be cooled is the positiveness of $J_{L}^{Q}$; therefore  $J_{L}$ is necessarily positive. This means that as long as the bias voltage is positive, electric power is being {\em invested} in the system, while when $V$ is negative (i.e., the electric current flows against the voltage) the device {\em  produces} electric power. Thus, the COP is given by $\eta_{a}$, Eq. (\ref{etaa}),  when the voltage is  positive and by $\eta_{b}$, Eq. (\ref{etab}), when it is negative.

Adding the expression for the heat current leaving the boson bath
\begin{align}
J^{Q}_{T}=\frac{\omega}{|e|}(|e|J^{}_{L})+
K^{}_{\rm PP}(1-\frac{T^{}_{R}}{T^{}_{T}})\ ,
\end{align}
we find that the COP's of our system are
\begin{align}
\eta^{}_{a}=\frac{B-\kappa^{}_{\rm P}(\frac{T^{}_{R}}{T^{}_{L}}-1)}{\frac{|e|V+\omega }{\overline{E}}B+\kappa^{}_{\rm PP}(1-\frac{T^{}_{R}}{T^{}_{T}})
}
\end{align}
for $V\geq 0$,  and
\begin{align}
\eta^{}_{b}=\frac{(1-\frac{|e|V}{\overline{E}})B
-\kappa^{}_{\rm P}(\frac{T^{}_{R}}{T^{}_{L}}-1)}{\frac{\omega }{\overline{E}}B+\kappa^{}_{\rm PP}(1-\frac{T^{}_{R}}{T^{}_{T}})}
\label{etabfi}
\end{align}
for $V\leq 0$.
Here we have introduced for brevity  the notation
\begin{align}
B=\frac{|e|V}{\overline{E}}+1-
\frac{T^{}_{R}}{T^{}_{L}}+\frac{\omega}{\overline{E}}(1-\frac{T^{}_{R}}{T^{}_{T}})\equiv \frac{e^2J^{}_L}{G^{}_{\rm in}\overline{E}}\ ,
\end{align}
and measured the phonon conductances in terms of the bare electronic heat conductance,
\begin{align}
\kappa^{}_{\rm P}=e^{2}K^{}_{\rm P}/(\overline{E}^{2}G^{}_{\rm in})\ ,\ \ \
\kappa^{}_{\rm PP}=e^{2}K^{}_{\rm PP}/(\overline{E}^{2}G^{}_{\rm in})\ .
\label{PHOC}
\end{align}
Note that the bias should exceed a certain threshold, $V_{c}$,  dictated by the working condition Eq. (\ref{wc}),
\begin{align}
\frac{|e|V^{}_{c}}{\overline{E}}=
\Big (\frac{T^{}_{R}}{T^{}_{L}}-1\Big )(1+\kappa^{}_{\rm P})
-\frac{\omega}{\overline{E}}\Big (1-\frac{T^{}_{R}}{T^{}_{T}}\Big )\ .
\label{vc}
\end{align}
As mentioned, 
 joint cooling and energy harvesting requires negative values of the bias voltage, and hence negative values of the threshold $V_{c}$;  a large enough  phonon conductance $K_{\rm P}$ (the actual value depends on the model parameters) will therefore prevent such a combined action.
\begin{figure}[htp]
\centering
\includegraphics[width=7cm]{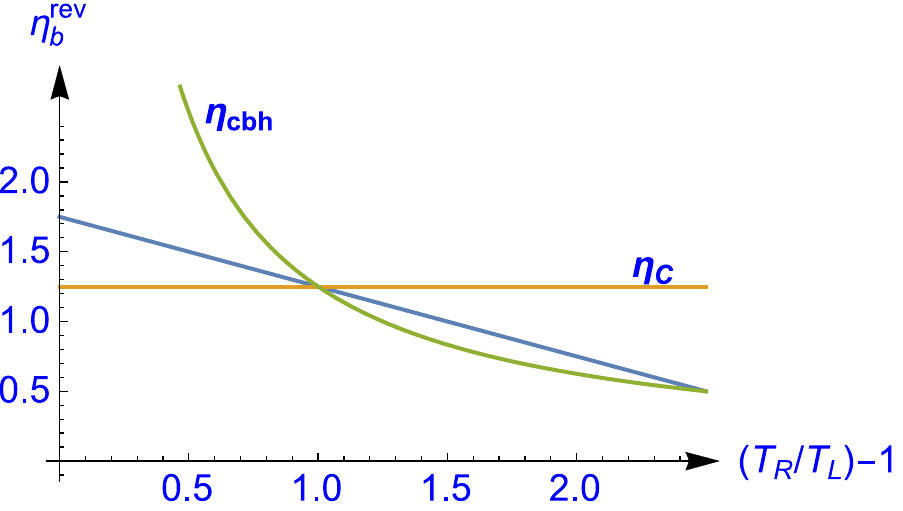}
\caption{(Color online) The slanted (blue) line is thermodynamic efficiency of a three-terminal junction, Eq. (\ref{theta}),  as a function of $(T_{R}/T_{L})-1$ for  $\kappa_{\rm P}=\kappa_{\rm PP}=0$, $(T^{}_{R}/T^{}_{L})-1=1.75$ and $\omega/\overline{E} =2$. The horizontal (orange) line is $\eta_{C}$, Eq. (\ref{Carnot}); the curved (green) line is $\eta_{\rm cbh}$, Eq. (\ref{etacbh}); the latter two represent  the bounds Eqs. (\ref{ine1}) and (\ref{ine2}).}
\label{fig3}
\end{figure}

It is illuminating at this point to consider the thermodynamic limit of the efficiency $\eta^{}_{ b}$, Eq. (\ref{etabfi}). Setting the thermal conductances to zero implies that the bias voltage is $V_{c}$, Eq. (\ref{vc}), and then
\begin{align}
\eta^{\rm rev}_{b}=\frac{\overline{E}}{\omega}
\Big [
1- \Big (\frac{T^{}_{R}}{T^{}_{L}}-1\Big )
+\frac{\omega}{\overline{E}}
\Big (1-\frac{T^{}_{R}}{T_{T}}\Big )\Big ]\ ,
\label{theta}
\end{align}
with $(T_{R}/T_{L})-1\leq (\omega/\overline{E})[1-(T_{R}/T_{T}) ]$ to ensure the negativeness of $V_{c}$.
We plot in Fig. \ref{fig3} this COP, together with the two bounds found in Sec. \ref{general} on it; indeed, the model-dependent $\eta^{\rm rev}_{b}$ of Eq. (\ref{theta}) lies in-between the bounds (\ref{ine1}) and (\ref{ine2}), which interchange their respective roles at $T_{R}/T_{L}=2$. Varying the model parameters does not change the figure qualitatively. As noted after Eq. (\ref{www}),  the reversible efficiency increases  upon generating  both cooling and electrical work (i.e.,  increasing  $w$) when $\eta_{\rm cbe}<1$ and decreases when $\eta_{\rm cbe}> 1$.

The figures below display the COP's $\eta_{a}$ and $\eta_{b}$ as functions of the bias voltage,  the former for positive values of the bias voltage and the latter for its negative values, as explained in Sec.  \ref{general}, see Eqs. (\ref{etaa}) and (\ref{etab}).  
In all figures the temperature difference between the electronic $L$ and $R$ terminals  is kept constant,  chosen to be $T_{R}/T_{L}=3/2$ and  yielding $\eta_{\rm cbe}=2$ for the Carnot coefficient of performance for refrigeration by investing work in a two-terminal setup [see Eq. (\ref{etacbe}) and the discussion in Sec. \ref{YI}].  The various curves in each figure are for different values of  $1-T_{R}/T_{T}$ (explicit values are given in the captions).
The parasitic phonon conductances [in dimensionless units, see Eqs. (\ref{PHOC})] of the electronic junction ($\kappa_{\rm P}$) and between the electronic system and the thermal terminal
($\kappa_{\rm PP}$) are taken to be equal for definiteness and decrease from 4 to 0.2, whereupon the COP increases.
In all three figures we see that the `working regime', i.e., the voltage range over which cooling is possible, increases with the
driving force of the thermal terminal $1-T_{R}/T_{T}$, namely the threshold $V_{c}$ decreases.

\begin{figure}[htp]
\centering
\includegraphics[width=7cm]{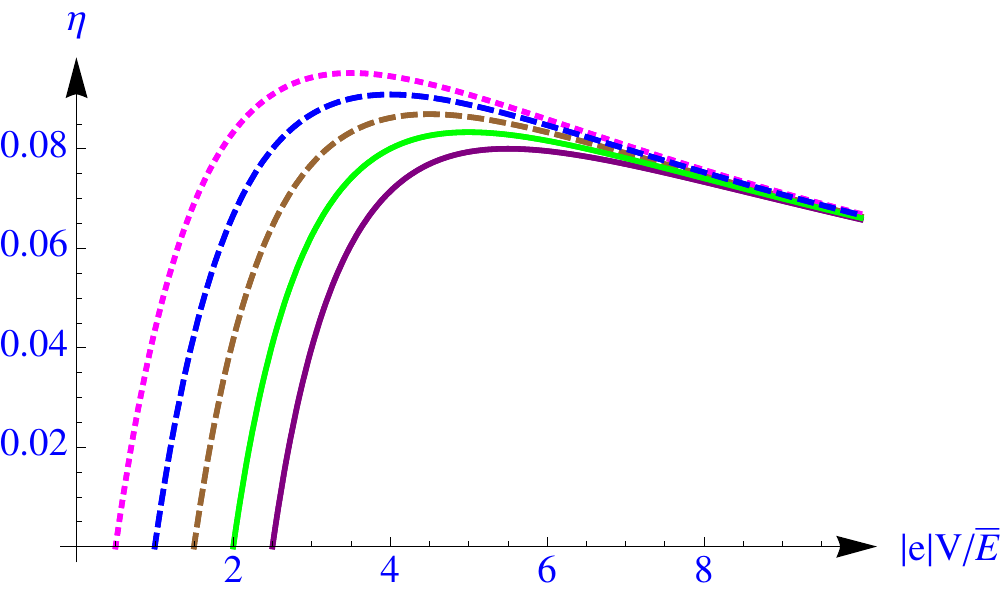}
\caption{(Color online) The efficiency of a three-terminal junction as a function of $|e|V/\overline{E}$ for various values of $1-T^{}_{R}/T^{}_{T}$, and $\kappa_{\rm P}=\kappa_{\rm PP}=4$, $(T^{}_{R}/T^{}_{L})-1=0.5$ and $\omega/\overline{E} =2$. The various curves in the upward direction are for $1-T_{R}/T_{T}=0$ [the lowest solid (purple) line],  $0.25,\ 0.5,\ 0.75$ and 1 [the upper dotted (magenta) curve].}
\label{c}
\end{figure}

\begin{figure}[htp]
\centering
\includegraphics[width=7cm]{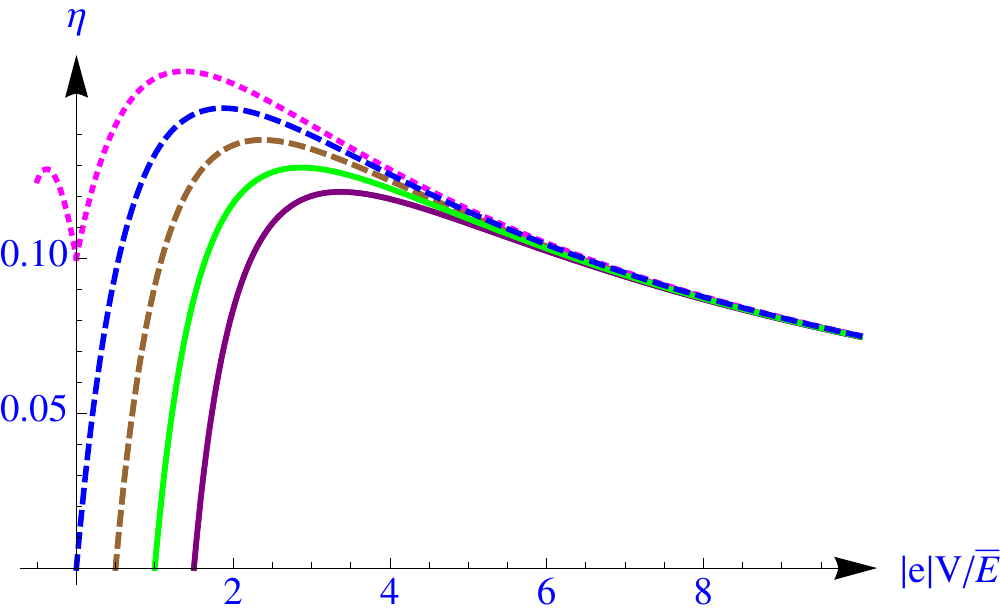}
\caption{(Color online) The efficiency of a three-terminal junction as a function of $|e|V/\overline{E}$ for various values of $1-T^{}_{R}/T^{}_{T}$, and $\kappa_{\rm P}=\kappa_{\rm PP}=2$, $(T^{}_{R}/T^{}_{L})-1=0.5$ and $\omega/\overline{E} =2$. The various curves in the upward direction are for $1-T_{R}/T_{T}=0$ 
[the lowest solid (purple) line],  $0.25,\ 0.5,\ 0.75$ and 1 [the upper dotted (magenta) curve].}
\label{b}
\end{figure}

\begin{figure}[htp]
\centering
\includegraphics[width=7cm]{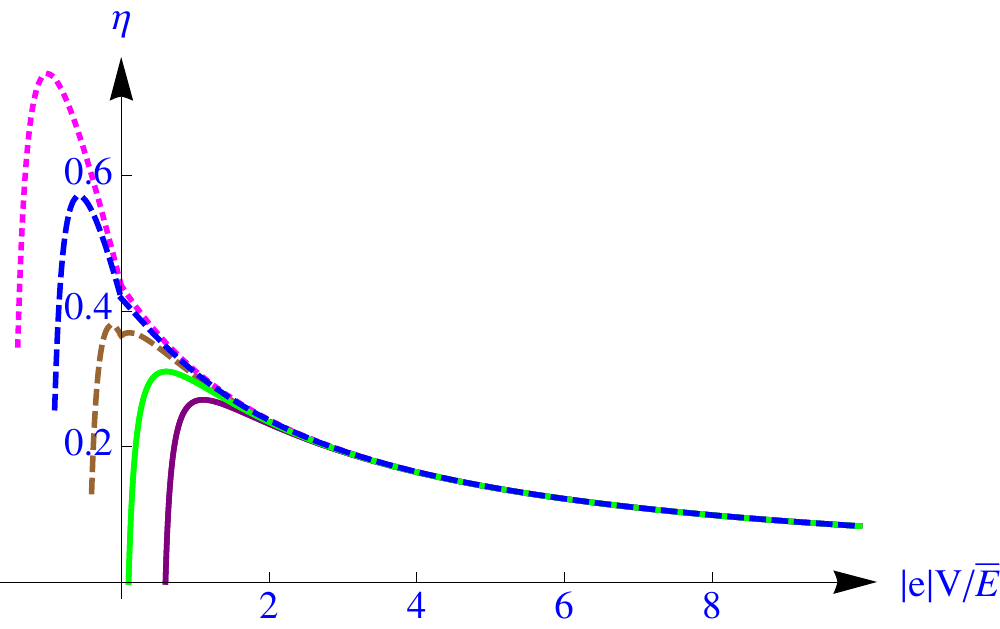}
\caption{(Color online) The efficiency of a three-terminal junction as a function of $|e|V/\overline{E}$ for various values of $1-T^{}_{R}/T^{}_{T}$, and  $\kappa_{\rm P}=\kappa_{\rm PP}=0.2$, $(T^{}_{R}/T^{}_{L})-1=0.5$ and $\omega/\overline{E} =2$. The various curves in the upward direction are for $1-T_{R}/T_{T}=0$ 
[the lowest solid (purple) line],  $0.25,\ 0.5,\ 0.75$ and 1 [the upper dotted (magenta) curve].}
\label{a}
\end{figure}

Figure \ref{c} is for $\kappa_{\rm P}=\kappa_{\rm PP}=4$; cooling exists only beyond a positive threshold of the bias voltage [see Eq. (\ref{vc})], so no joint cooling and energy harvesting is possible.
However, that is already possible for $\kappa_{\rm P}=\kappa_{\rm PP}=2$, as shown in Fig. \ref{b}. Very interestingly, the COP increases with decreasing $V<0$ and has a small peak, which is yet smaller than the one for $V>0$. What is, seemingly, most surprising is that for even smaller parasitics, e.g.,  $\kappa_{\rm P}=\kappa_{\rm PP} = .2$ (Fig. \ref{a}), still away from the reversible limit, 
the values of the COP at $V<0$ can be substantially larger than those for $V>0$. This means that harvesting energy may actually {\em increase the COP compared to investing energy}, a truly remarkable property of the three-terminal setup!

\begin{figure}
\includegraphics[width=7cm]{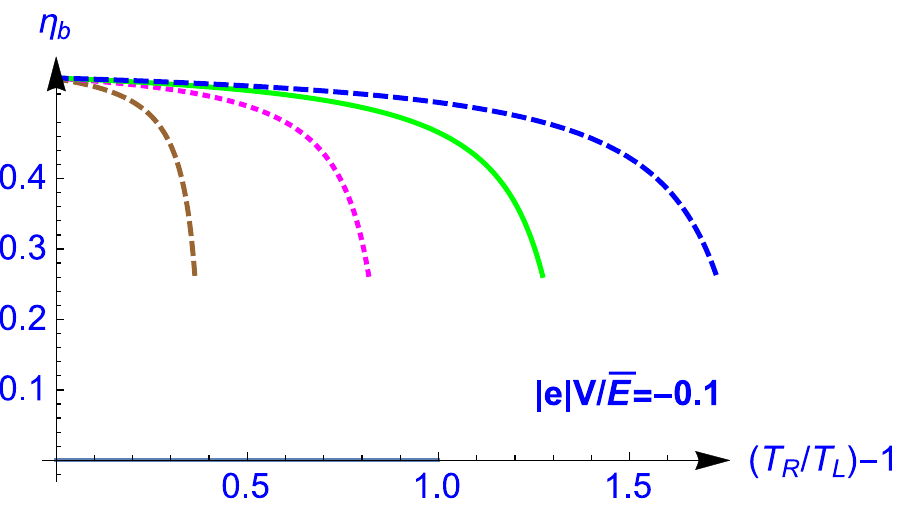}
\includegraphics[width=7cm]{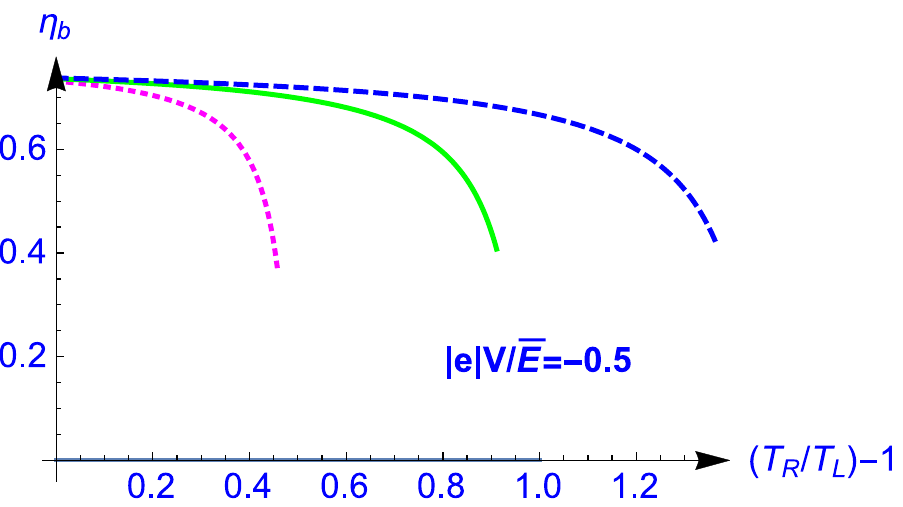}
\includegraphics[width=7cm]{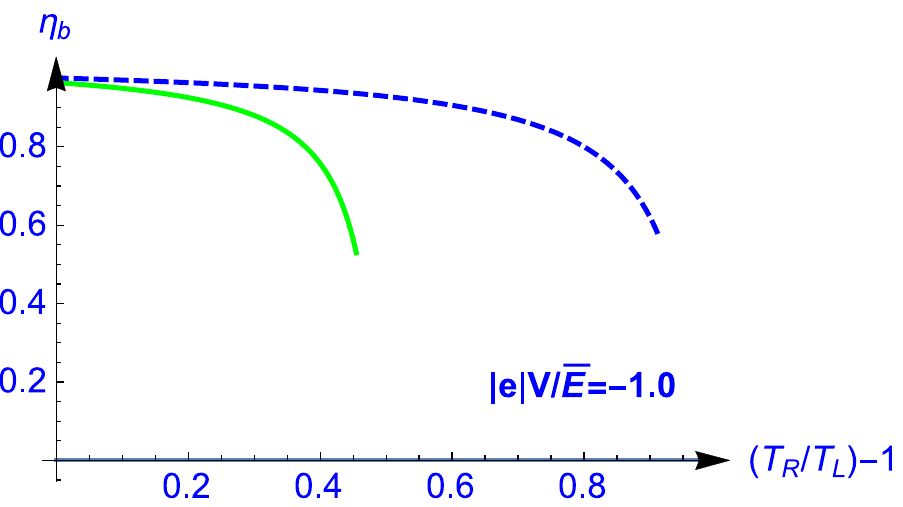}\caption{(Color online) The efficiency $\eta _{b}$ 
as a function of the temperature difference across the electronic system, for various values of the temperature difference between the thermal reservoir and the electrons, $1-(T_{R}/T_{T})=0.25$, the dashed (brown) curve, $=0.5$,  the dotted (magenta) curve, $=0.75, $ the solid (green) line and $=1.$,  the dashed (blue) curve. The upper panel is for $|e|V/\overline{E}=-0.1$, the middle one is for
$|e|V/\overline{E}=-0.5$, and the lower panel is for
$|e|V/\overline{E}=-1$. The working condition Eq. (\ref{vc}) is not fulfilled for all values of $1-(T_{R}/T_{T})=0.25$ and therefore some of the curves are missing in the two lower panels.  Here
$\kappa_{\rm P}=\kappa_{\rm PP}=0.05$,  and $\omega/\overline{E} =2$.
}
\label{C}
\end{figure}

The tendency of the efficiency to increase with the absolute value of the voltage for small negative values of it is
exemplified by Figs. \ref{C} which display $\eta_{b}$, Eq. (\ref{etabfi}), as a function of $(T_{L}/T_{R})-1$ for negative values of the voltage. It is seen that indeed $\eta_{b}$ increases as the absolute value of $V$ increases. However, because of the  condition Eq. (\ref{vc}) necessary for the device to operate, the range of temperature differences between the thermal terminal and the electrons is reduced as well. The various curves in each of the panels in Figs. \ref{C} correspond to different values of
$1-T_{R}/T_{T}$. As $|V|$ is increased, the working condition Eq. (\ref{vc}) allows only for  $T_{R}-$values which are  closer and closer to $T_{T}$.

\section{Summary and conclusions}

\label{SUM}

We have analyzed the efficiency of two cooling processes possible in a mixed three-terminal thermoelectric junction, comprising
two electronic terminals that interchange energy with a third, thermal contact. For concreteness, we have concentrated mainly on the possibility to cool one of the electronic terminals, either by investing  thermal energy extracted from the thermal terminal {\em and} electric power supplied by a bias voltage on the electrons, or by exploiting the thermal energy of the boson bath to cool the electronic terminal
{\em
 and to  produce } electric power.  (Cooling of the thermal contact has been mentioned in Sec. \ref{YI}.)
We have found that  one advantage of the three-terminal setup compared with the two-terminal one is the increase in the working regime of the device. In our case, this is manifested by the extended range of bias-voltage values for which cooling is possible. This increase is dominated by the parasitic phonon conductance; in our  case it is the phonon conductance, $K_{\rm P}$,  of the electronic junction, which should not be too large.  With a suitable choice  of $K_{\rm P}$ the threshold of the voltage becomes negative, and then the three-terminal setup cools and at the same time, produces electric power.  In our model system, we find that the COP for this  dual action can be {\em enhanced} compared with its value (for the same parameters) obtained when the  device works just as a refrigerator.   This enhancement, necessitating not-too-large parasitic thermal conductances,  should become huge when the cooling is for a small cooling temperature increment, $T_R - T_L << T_L$.

\begin{acknowledgments}
This work was supported by the Israeli Science Foundation (ISF) and the US-Israel Binational Science Foundation (BSF).  We thank J-H Jiang, J-L Pichard, D. Shahar, G. C. Tewari and G. Zeltzer  for discussions on related questions.
OEW and AA thank FAPESP grant 2011/11973-4 for funding their visit to ICTP-SAIFR during  February-March 2014,  where part of this work was done.

\end{acknowledgments}


\begin{thebibliography}{999}
\bibitem{com1}
See Ref. \onlinecite{sign}: as  the left electronic terminal is to be cooled, $J_{L}^{Q}$  is positive.  This is the `working condition'  of the device.
\bibitem{com2}
Phonon-drag contributions may be ignored in this localized-state electronic junction.
\end{thebibliography}
\end{document}